\documentclass[english,pra,aps,superscriptaddress,longbibliography,notitlepage,twocolumn]{revtex4-1}
\usepackage[T1]{fontenc}
\usepackage[latin9]{inputenc}
\usepackage{array}
\usepackage{booktabs}
\usepackage{mathrsfs}
\usepackage{multirow}
\usepackage{amsmath}
\usepackage{graphicx}

\makeatletter

\providecommand{\tabularnewline}{\\}

\usepackage[colorlinks,citecolor=blue,linkcolor=blue]{hyperref}
\usepackage{booktabs}

\usepackage{array}
\usepackage[table,xcdraw]{xcolor}
\usepackage{multirow}

\makeatother

\usepackage{babel}
\begin{document}
\title{Simulating finite-time quantum isothermal processes with generic superconducting
quantum circuit}
\date{\today}
\author{Jin-Fu Chen}
\address{Beijing Computational Science Research Center, Beijing 100193, China}
\address{Graduate School of China Academy of Engineering Physics, No. 10 Xibeiwang
East Road, Haidian District, Beijing, 100193, China}
\author{Ying Li}
\address{Graduate School of China Academy of Engineering Physics, No. 10 Xibeiwang
East Road, Haidian District, Beijing, 100193, China}
\author{Hui Dong}
\email{hdong@gscaep.ac.cn}

\address{Graduate School of China Academy of Engineering Physics, No. 10 Xibeiwang
East Road, Haidian District, Beijing, 100193, China}
\begin{abstract}
The finite-time isothermal process is fundamental in quantum thermodynamics
yet complicated with combination of changing control parameters and
the interaction with the thermal bath. Such complexity prevents the
direct application of the traditional thermodynamics measurement of
the relevant quantities. In this paper, we provide a discrete-step
method to separate the work done and the heat exchange in the isothermal
process by decomposing the process into piecewise adiabatic and isochoric
processes. The piecewise control scheme makes it possible to simulate
the whole process on a generic quantum computer, which provides a
new platform to experimentally study quantum thermodynamics. We implement
the simulation on ibmqx2 to show the $\mathrm{\mathcal{C}/\tau}$
scaling of the extra work in the finite-time isothermal process.
\end{abstract}
\maketitle

\section{Introduction}

Quantum thermodynamics, originally believed as an extension of classical
thermodynamics, has sharpen our understanding toward the fundamental
aspects of thermodynamics \citep{Campisi_2011,Esposito_2009,Maruyama2009,Linden2010PhysRevLett105_130401,Strasberg_2017,Vinjanampathy_2016}.
Along with the theoretical progresses, experimental tests and validation
of the underline principles are relevant in the realm. Simulation
of the quantum thermodynamic phenomena \citep{Feynman1982,Lloyd1996,Georgescu2014,Sandholzer2019},
as one of the experimental efforts, has been intensively explored
with specifically designed system for specific purpose, e.g., the
trapped ion for testing the Jarzynski equation \citep{An2014}, the
BEC system for quantum work extraction \citep{Deng2015}, and the
superconducting qubit for the shortcut to adiabatic \citep{Wang2019}.
These designed systems often have limited functions to test specific
quantum thermodynamic properties. Simulation with generic quantum
computing system shall offer a universal system to demonstrate essential
quantum thermodynamic phenomena.

Yet, simulation of quantum thermodynamics with the universal quantum
computer remains a challenge mainly due to the lack of the flexibility
to physically tune control parameters and the difficulty to measure
the work extraction. In quantum thermodynamics, the work extraction,
as a fundamental quantity \citep{Alicki1979,Quan_2007,Su2018}, requires
the ability to tune the control parameters. Such requirement is achievable
in the specifically designed system, e.g., the force shift in the
trapped ion \citep{Blatt2012,An2014}, the trapped frequency in BEC
\citep{Deng2015} and the transition frequency in the superconducting
system \citep{Wang2019}. However, in the universal optimized quantum
computer, e. g. IBM quantum computer (ibmqx2), the operations are
limited for the users and such tuning is unfortunately not available.
Additional problem is the measurement of the work extraction, which
can be easily measured in classical thermodynamics by recoding the
control parameters and measuring the conjugate quantities. Such measurement
remains a challenge in the quantum region \citep{Talkner2016}.

In this paper, a proposal is given and experimentally demonstrated
to simulate the basic non-equilibrium thermodynamics process with
quantum computer by introducing a virtual way to tune the control
parameters, i.e. without physically tuning the parameters. The dynamics
are realized using quantum gates according to the parameter changes.
As a demonstration, we realize the simulation on ibmqx2 \citep{ibm}
for the isothermal process, which is a fundamental process for quantum
heat engine cycle, yet complicated with both the changing control
parameters and the interaction with a thermal bath. To implement the
simulation using, we consider the discrete-step method to approach
the quantum isothermal process: the isothermal process is divided
into series of elementary processes, each consisting of the adiabatic
process and the isochoric process. In the adiabatic process, the parameter
adjusting is performed virtually and the unitary evolution is implemented
with quantum gates. In the isochoric process, the effect of the bath
contact is performed with quantum channel simulation \citep{Ruskai2002,Nielsen2009,Fisher2012,Wang2013,Lu2017,Hu2018}
with the assist of ancillary qubits for quantum dynamics simulation
in open quantum systems\citep{Wang2011,Uzdin2019,Su2020}. With this
approach, we achieve the simulation of the isothermal process on the
universal quantum computing system without flexibility to directly
adjust physical parameters such as the energy levels.

The simulation with universal quantum computer brings clear advantages
in our proposal. First, \textit{the arbitrary change of the control
parameters} is archived by the virtual tuning via the simulation of
corresponding dynamics, avoiding the difficulty in tuning the actual
physical system. In turn, the parameters can be controlled to follow
an arbitrary designed functions. Second, we can realize \textit{the
immediate change of environment parameters,} such as the temperature.
The effect of the bath is reflected through the state of the auxiliary
qubit, which can be controlled flexibly using quantum gates.

This paper is organized as follows. In Sec. \ref{sec:Rectangle-integration-approach-t},
we introduce the discrete-step method to approach the quantum isothermal
process. The isothermal process is approached with series of the adiabatic
and the isochoric processes. In Sec. \ref{sec:Simulation-with-quantum},
we present circuit realization of one elementary process. Two methods,
the hybrid simulation and the fully quantum simulation, are designed
to simulate the quantum thermodynamic process using quantum circuits.
In Sec. \ref{sec:Testing--}, we show an application of our simulation
scheme. The ibmqx2 simulation results of the discrete isothermal process
are compared with the numerical results. The conclusion is given in
Sec. \ref{sec:Conclusion}.

\section{discrete-step method to quantum isothermal process\label{sec:Rectangle-integration-approach-t}}

In quantum thermodynamics, the generic evolution of the concerned
system can be considered as a quantum evolution with the changing
Hamiltonian while in contact with a thermal bath. The interplay between
work and heat during the general process has limited the characterization
of the quantum thermodynamic cycle on the microscopic level, where
the classical method to measure the work, via force and distance,
is not applicable \citep{Talkner2016}. In the limit of timescale
of tuning the control parameter far larger than the thermal bath response
time, the evolution is the thermodynamic adiabatic process, where
the heat exchange with the bath is neglected and the internal energy
changes due to the work done through the changing control field. Another
extreme case with unchanged control parameters is known as the isochoric
process, where the internal energy changes solely by the heat exchange
with the thermal bath. Therefore, the work and heat can be separated
clearly in the adiabatic and isochoric process, and be obtained directly
by measuring the internal energy change.

\begin{figure}
\textbf{\includegraphics[width=7cm]{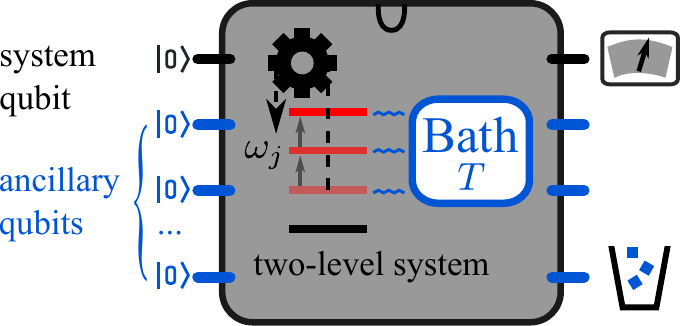}}

\caption{Isothermal process of a two-level system is simulated on the superconducting
quantum computer with the discrete-step method. The discrete-step
method utilizes series of adiabatic processes and the isochoric processes
to approach the finite-time isothermal process. In the adiabatic process,
the energy of the two-level system is tuned and the interaction between
the system and the bath is switched off. In the isochoric process,
the interaction is switched on with the fixed excited-state energy
$\omega_{j}$. In the simulation, one qubit is used to represent the
simulated two-level system, and the ancillary qubits play the role
of the thermal bath at the temperature $T$. After implementing the
quantum circuit, the system qubit is measured to obtain the internal
energy of the simulated two-level system. \label{fig:(a)-Rectangle-Integration-approa}}
\end{figure}

One method to simulate the generic quantum thermodynamic process while
keeping the separation of the work and heat, is to use the discrete-step
method with a series of adiabatic and isochoric processes \citep{Quan_2008,Ma2018}.
In Fig. \ref{fig:(a)-Rectangle-Integration-approa}, the discrete-step
method is illustrated with the minimal model, a two-level system with
the energy spacing $\omega(t)$ between the ground state $\left|g\right\rangle $
and the excited state $\left|e\right\rangle $. Such two-level system
can be physically realized with a qubit, which is a elementary unit
of the quantum computer. For the clarity of the later discussion,
we use the term ``two-level system'' to denote the simulated system,
and the term ``qubit'' as the simulation system hereafter without
specific mention. The state of the two-level system is represented
by the density matrix $\rho_{s}(t)$ of the system qubit, and the
effect of the thermal bath is simulated by the assist of some ancillary
qubits. Initially, the system qubit is prepared to the thermal state
$\rho_{s}(0)$ with the corresponding energy $\omega_{0}$ and the
temperature $T$. The evolution of the open quantum system can be
implemented with single-qubit and two-qubit quantum gates on the superconducting
quantum computer. The internal energy of the two-level system is determined
by $E(t)=\omega(t)p_{e}(t)$, with the population of the excited state
$p_{e}(t)=\left\langle e\right|\rho_{s}(t)\left|e\right\rangle $.

For the system to be simulated, we use the discrete-step method to
approach the finite-time isothermal process for the two-level system.
The discrete isothermal process contains $N$ steps of elementary
processes with the total time consuming $\tau+\tau_{\mathrm{adi}}$,
where $\tau$ ($\tau_{\mathrm{adi}}$) denotes the time consuming
in the isochoric (adiabatic) process. Each elementary process is composed
by an adiabatic process and an isochoric process. We set equal time
consuming for every elementary process $\delta\tau=(\tau+\tau_{\mathrm{adi}})/N$,
with $\tau/N$ ($\tau_{\mathrm{adi}}/N$) for each isochoric (adiabatic)
process.

In the adiabatic process, the system is isolated from the bath and
evolves under the time-dependent Hamiltonian. Such process is described
by a unitary evolution with the control time as $\tau_{\mathrm{adi}}/N$.
The performed work is determined by the change of the internal energy
at the initial and the final time. For a generic adiabatic process,
the unitary evolution of the system can be simulated with the single-qubit
gate acted on the system qubit. In this paper, we consider the adiabatic
process as the instant quench with zero time consuming $\tau_{\mathrm{adi}}=0$,
occurred at time $t_{j-1}=(j-1)\delta\tau,\,j=1,2,...,N.$ As the
result of the quench, the energy of the excited state is shifted from
$\omega_{j-1}$ to $\omega_{j}$, while the density matrix $\rho_{s}(t_{j-1})$
remains unchanged after the quench. At the initial time $t_{0}=0$,
the energy is quenched from $\omega_{0}$ to $\omega_{1}$ after the
initial state preparation. The performed work for the quench at time
$t_{j-1}$ reads
\begin{equation}
W_{j}=(\omega_{j}-\omega_{j-1})p_{e}(t_{j-1}).
\end{equation}
To obtain the performed work for the simulated finite-time quantum
isothermal process, we only need to obtain the diagonal element $p_{e}(t_{j-1})=\left\langle e\right|\rho_{s}(t_{j-1})\left|e\right\rangle $
via measuring the state population of the system qubit at the beginning
of each isochoric process.

In the isochoric process of the $j$-th elementary process ($t_{j-1}<t\leq t_{j}$),
the two-level system is brought into contact with the thermal bath
at the temperature $T$. The evolution is given by the master equation

\begin{align}
\dot{\rho}_{s} & =-i[H_{j},\rho_{s}]+\gamma_{0}N_{j}\mathscr{L}(\sigma_{+})[\rho_{s}]\nonumber \\
 & +\gamma_{0}(N_{j}+1)\mathscr{L}(\sigma_{-})[\rho_{s}],\label{eq:master1}
\end{align}
with
\begin{equation}
\mathscr{L}(\sigma)[\rho_{s}]=\rho_{s}\sigma^{\dagger}-\frac{1}{2}\sigma^{\dagger}\sigma\rho_{s}-\frac{1}{2}\rho_{s}\sigma^{\dagger}\sigma.
\end{equation}
Here, $H_{j}=\omega_{j}\left|e\right\rangle \left\langle e\right|$
is the Hamiltonian of the system during the period $t_{j-1}<t\leq t_{j}$,
$N_{j}=1/[\exp(\beta\omega_{j})-1]$ is the average photon number
with the inverse temperature $\beta=1/(k_{\mathrm{B}}T)$, and $\sigma_{+}=\left|e\right\rangle \left\langle g\right|$
($\sigma_{-}=\left|g\right\rangle \left\langle e\right|$) is the
transition operator. In this process, the change of the internal energy
is induced by the heat exchange with the thermal bath, and no work
is performed. During the whole discrete isothermal process, the work
is only performed at the time $t_{j}$.

\section{Simulation with quantum circuits\label{sec:Simulation-with-quantum}}

In this section, we first show the simulation of one elementary process
with quantum gates in the circuit. The simulation is formulated for
the adiabatic and the isochoric processes as follows.

\textbf{\begin{flushleft}I. Adiabatic process: \end{flushleft}}

In the superconducting quantum computer, e.g., IBM Q system, the tuning
of the physical energy levels of qubits is unavailable for the users.
Usually, the parameters of the system are fixed at the optimal values
to possibly reduce noise and error induced by decoherence and imperfect
control. The Hamiltonian of the simulated two-level system is modulated
as $H(t)=\omega(t)\left|e\right\rangle \left\langle e\right|$, where

\begin{equation}
\omega(t)=\omega_{j},\quad t_{j-1}<t\leq t_{j}\:\mathrm{with}\;j=1,2...,N.\label{eq:energy modulate-1}
\end{equation}
We will show that the shifted energy $\omega(t)$ of the simulated
two-level system only affects the transition rate induced by the thermal
bath. In the simulation, the thermal transition is simulated through
the quantum channel simulation with the assist of ancillary qubits,
and the transition rate can be flexibly tuned by the single-qubit
gates on the ancillary qubits. Therefore, we do not have to physically
tune any parameters of the quantum computer, and just algorithmically
modulate the simulated thermal transition instead. We propose a virtual
tuning of the energy levels with details explained as follows.

In the virtual process, we need to simulate the unitary evolution
of the adiabatic process with single-qubit gates on the system qubit.
For the adiabatic process, i.e. the quench, the state of the system
does not evolve in the short period. We don't have to do anything
with the physical system, and just pretend that the energy of the
simulated system is tuned from $\omega_{i}$ to $\omega_{j+1}$ in
the $j$-th adiabatic process. This virtual tuning of the energy is
reflected by the modulation of the thermal transition rate in the
simulation of the isochoric process.

\textbf{\begin{flushleft}II. Isochoric process: \end{flushleft}}

\begin{figure*}
\includegraphics[scale=0.8]{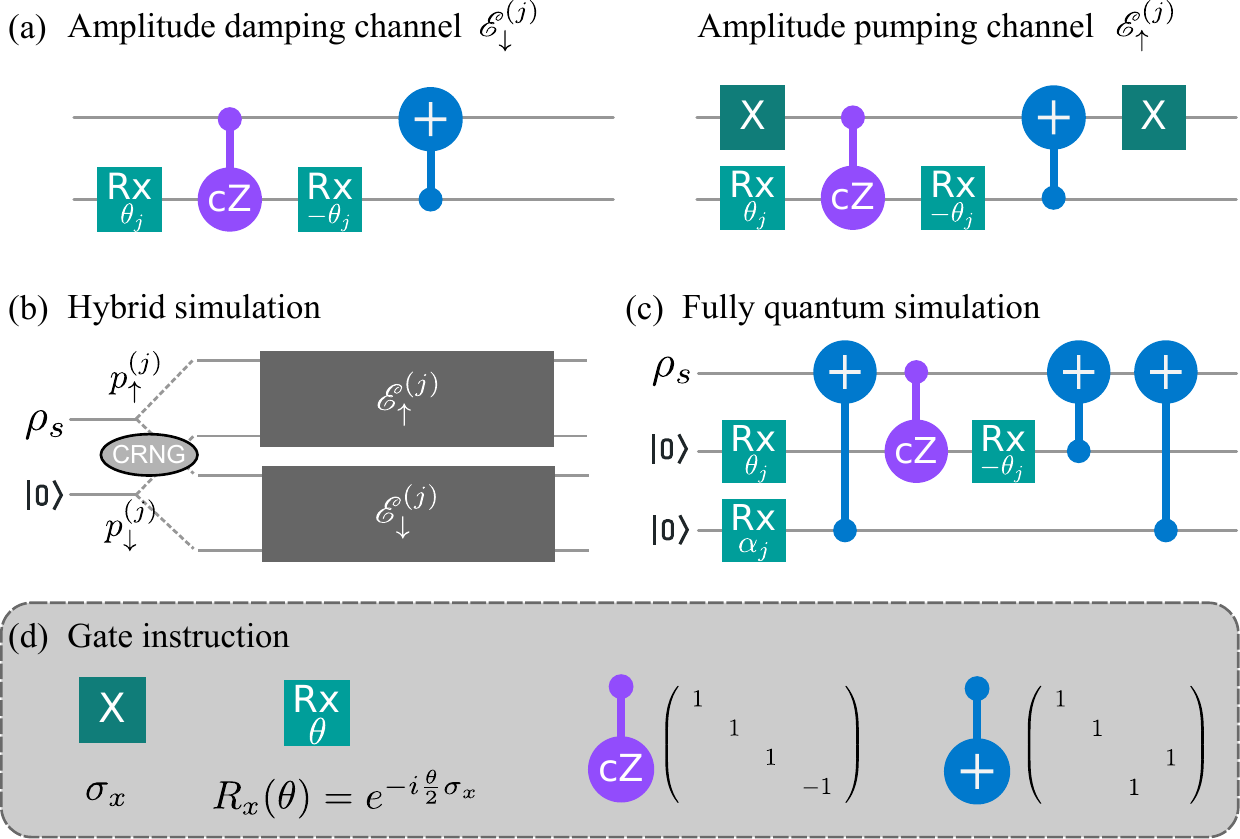}

\caption{The quantum circuits to simulate the isochoric process in one elementary
process. (a) The quantum circuits of the amplitude damping (pumping)
channel $\mathscr{E}_{\downarrow}^{(j)}$ ($\mathscr{E}_{\uparrow}^{(j)}$)
in the hybrid simulation. (b) The quantum circuits for the hybrid
simulation. The selection of the two sub-channels is realized using
the classical random number generator. (c) The quantum circuit for
the fully quantum simulation. The selection of the two sub-channels
is assisted by another ancillary qubit. (d) Instruction of gate in
the current simulation.}

\label{fig:elementarycircuit}
\end{figure*}

The simulation of the isochoric process is to simulate the dynamics
governed by the master equation in Eq. (\ref{eq:master1}). The dynamical
evolution of the isochoric process can be simulated with the generalized
amplitude damping channel (GADC)

\begin{equation}
\rho_{s}(t_{j})=\mathscr{E}_{\mathrm{GAD}}^{(j)}[e^{-iH_{j}\delta\tau}\rho_{s}(t_{j-1})e^{iH_{j}\delta\tau}],\label{eq:generalized}
\end{equation}
where $\mathscr{E}_{\mathrm{GAD}}^{(j)}=p_{\downarrow}^{(j)}\mathscr{E}_{\downarrow}^{(j)}+p_{\uparrow}^{(j)}\mathscr{E}_{\uparrow}^{(j)}$
is divided into two sub-channels, the amplitude damping channel
\begin{equation}
\mathscr{E}_{\downarrow}^{(j)}[\rho_{s}]=M_{0}^{(j)}\rho_{s}M_{0}^{(j)\dagger}+M_{1}^{(j)}\rho_{s}M_{1}^{(j)\dagger},
\end{equation}
and the amplitude pumping channel
\begin{equation}
\mathscr{E}_{\uparrow}^{(j)}[\rho_{s}]=M_{2}^{(j)}\rho_{s}M_{2}^{(j)\dagger}+M_{3}^{(j)}\rho_{s}M_{3}^{(j)\dagger}.
\end{equation}
The corresponding Kraus operators are $M_{0}^{(j)}=\cos\theta_{j}\left|e\right\rangle \left\langle e\right|+\left|g\right\rangle \left\langle g\right|$,
$M_{1}^{(j)}=\sigma_{-}\sin\theta_{j}$, $M_{2}^{(j)}=\left|e\right\rangle \left\langle e\right|+\cos\theta_{j}\left|g\right\rangle \left\langle g\right|$
and $M_{3}^{(j)}=\sigma_{+}\sin\theta_{j}$. The coefficient $p_{\uparrow}^{(j)}=1/[\exp(\beta\omega_{j})+1]$
($p_{\downarrow}^{(j)}=1-p_{\uparrow}^{(j)}$) shows the probability
of excitation (de-excitation) of the two-level induced by the thermal
bath. The evolution time of the $j$-th elementary process is encoded
in the control parameter $\theta_{j}$ via the relation
\begin{equation}
\cos\theta_{j}=\exp(-\frac{\gamma_{0}\delta\tau}{2}\frac{e^{\beta\omega_{j}}+1}{e^{\beta\omega_{j}}-1}).\label{eq:theta}
\end{equation}
The quasi-static discrete isothermal process with infinite evolution
time can be realized by setting $\theta_{j}=\pi/2$, where the system
reaches thermal equilibrium at the end of each isochoric process.

For the initial thermal state $\rho_{s}(0)=e^{-\beta H(0)}/\mathrm{Tr}(e^{-\beta H(0)})$,
the off-diagonal term remains zero throughout the whole discrete isothermal
process in the current control scheme. In this situation, the evolution
by Eq. (\ref{eq:generalized}) is simplified as 
\begin{equation}
\rho_{s}(t_{j})=\mathscr{E}_{\mathrm{GAD}}^{(j)}[\rho_{s}(t_{j-1})].\label{eq:gadsimple}
\end{equation}
For an initial state with non-zero non-diagonal term, the coherence
does not affect the diagonal term of the density matrix. This comes
from the fact that the diagonal term and the non-diagonal term of
the density matrix satisfy respective differential equations by Eq.
(\ref{eq:master1}).

Figure \ref{fig:elementarycircuit} shows the quantum circuit to simulate
the isochoric process. The two sub-channels $\mathscr{E}_{\downarrow}^{(j)}$
and $\mathscr{E}_{\uparrow}^{(j)}$ are realized with an ancillary
qubit initially prepared in the ground state \citep{Hu2018}. The
circuits for these two sub-channels are illustrated in Fig. \ref{fig:elementarycircuit}(a).
The meaning of each gate is explained at the bottom of Fig. \ref{fig:elementarycircuit}.
Such simulation circuits are extensively studied in the field of quantum
computing and quantum information that we will not explain the setup
in detail \citep{Nielsen2009}.

For the probabilities $p_{\downarrow}^{(j)}$ and $p_{\uparrow}^{(j)}$
in selecting the two sub-channels, we design two methods, the hybrid
simulation and the fully quantum simulation, to achieve the random
selection, as shown in Fig. \ref{fig:elementarycircuit}. (b) and
(c). The former uses one ancillary qubit for each elementary process
with the assist of a classical random number generator (CRNG), which
can be realized using a quantum computer with fewer qubits. The latter
utilizes fully quantum circuit with two ancillary qubits for each
elementary process, which can be realized using a quantum computer
with adequate qubits.

\begin{table*}[ht]
\begin{tabular}{>{\raggedright}p{0.14\linewidth}>{\raggedright}p{0.25\linewidth}>{\raggedright\arraybackslash}p{0.26\linewidth}>{\raggedright\arraybackslash}p{0.26\linewidth}}
\toprule 
\multirow{1}{0.14\linewidth}{} & To be simulated: & \multicolumn{2}{c}{Simulation}\tabularnewline
\cmidrule{3-4} \cmidrule{4-4} 
 & Discrete isothermal process & Hybrid simulation with CRNG & Fully quantum simulation\tabularnewline
\midrule 
\rowcolor[HTML]{EFEFEF} Adibatic

process & $U[R(t)]$, $t\in[\tau_{i},\tau_{i+1}]$ & \multicolumn{2}{p{0.54\linewidth}}{The unitary evolution is realized with the virtual tuning on the system
Hamiltonian.}\tabularnewline
Isochoric

process & System relaxtion in Eq. \ref{eq:master1} & Generalized amplitude damping channel ($\mathscr{E}_{\mathrm{GAD}}^{(j)}$)
with the classical random number generation & Generalized amplitude damping channel ($\mathscr{E}_{\mathrm{GAD}}^{(j)}$)
with an additional qubit at the state $\cos(\alpha_{j}/2)\left|0\right\rangle +i\sin(\alpha_{j}/2)\left|1\right\rangle $\tabularnewline
\rowcolor[HTML]{EFEFEF} Parameters & Time Duration: $\delta\tau=t_{j+1}-t_{j}$

Temperature: $T$ & $\cos\theta_{j}=\exp(-\frac{\gamma_{0}\delta\tau}{2}\frac{e^{\beta\omega_{j}}+1}{e^{\beta\omega_{j}}-1})$ & $\cos\theta_{j}=\exp(-\frac{\gamma_{0}\delta\tau}{2}\frac{e^{\beta\omega_{j}}+1}{e^{\beta\omega_{j}}-1})$
$\cos(\alpha_{j}/2)=[p_{\downarrow}^{(j)}]^{1/2}$\tabularnewline
\bottomrule
\end{tabular}\caption{The discrete isothermal process to be simulated and the two simulation
methods, the hybrid simulation and the fully quantum simulation.}
\end{table*}

\textbf{\begin{flushleft}1. Hybrid simulation of isochoric process
with classical random number generator (CRNG)\end{flushleft}}

With the limited number of qubits, it is desirable to save the unnecessary
usage of qubits. To simulate the quantum channel for the system qubit,
one ancillary qubit is inevitably used to simulate the non-unitary
evolution of the open quantum system \citep{Wang2013}. In this design,
one qubit is used to represent the two-level system, and one more
qubit is needed for adding one elementary process. Therefore, it requires
$N+1$ qubits to simulate the $N$-step isothermal process.

In the hybrid simulation, the CRNG is used to select the sub-channel
$\mathcal{O}_{j}^{[l]}=\mathscr{E}_{\uparrow}^{(j)}$ or $\mathscr{E}_{\downarrow}^{(j)}$
for the isochoric process in the $j$-th elementary process, as shown
in Fig. \ref{fig:elementarycircuit}(b). $l$ denotes the $l$-th
simulation of the discrete isothermal process. For each isochoric
process, the CRNG generates a random number $r_{j}^{[l]}\in[0,1]$
with uniform distribution. The sub-channel $\mathcal{O}_{j}^{[l]}$
is selected as $\mathscr{E}_{\downarrow}^{(j)}$ ($\mathscr{E}_{\uparrow}^{(j)}$
) when the random number satisfies $r_{j}^{[l]}\leq p_{\downarrow}^{(j)}$
($r_{j}^{[l]}>p_{\downarrow}^{(j)}$).

\textbf{\begin{flushleft}2. Fully quantum simulation of isochoric
process \end{flushleft}}

For the system with adequate qubits, the selection of the two sub-channels
can be achieved with an additional ancillary qubit, which encodes
the classical probability, as shown in Fig. \ref{fig:elementarycircuit}(c).
In each step, one more ancillary qubit is used, prepared to the super-position
state $\cos(\alpha_{j}/2)\left|0\right\rangle +i\sin(\alpha_{j}/2)\left|1\right\rangle $
through the $R_{x}(\alpha_{j})$ gate with $\cos(\alpha_{j}/2)=[p_{\downarrow}^{(j)}]^{1/2}$.
Since two ancillary qubits are needed for adding one elementary process,
this method requires $2N+1$ qubits to simulate the $N$-step isothermal
process.

In Table I, we summarize the simulation procedure for the adiabatic
and the isochoric processes. The correspondence between the parameters
in the simulated system and the simulation is listed in the table.

In the current simulation scheme, we solve the problem of separating
work and heat. The unitary evolution of the adiabatic process requires
isolation from the environment, while the isochoric process needs
the contact with environment. Switching on and off the interaction
with the thermal bath is complicated and requires enormous efforts,
especially in quantum region for a microscopic system. Fortunately,
the design of the quantum computer with long coherent time can easily
meet the first requirement of the isolation from the environment.
The simulation of quantum channel is designed to simulate the effect
of the environment. The advantage of quantum channel simulation over
the real coupling to the environment is its flexibility to control
parameters, e.g., the temperature, the coupling strength, et. al.

\begin{figure*}
\includegraphics[scale=0.8]{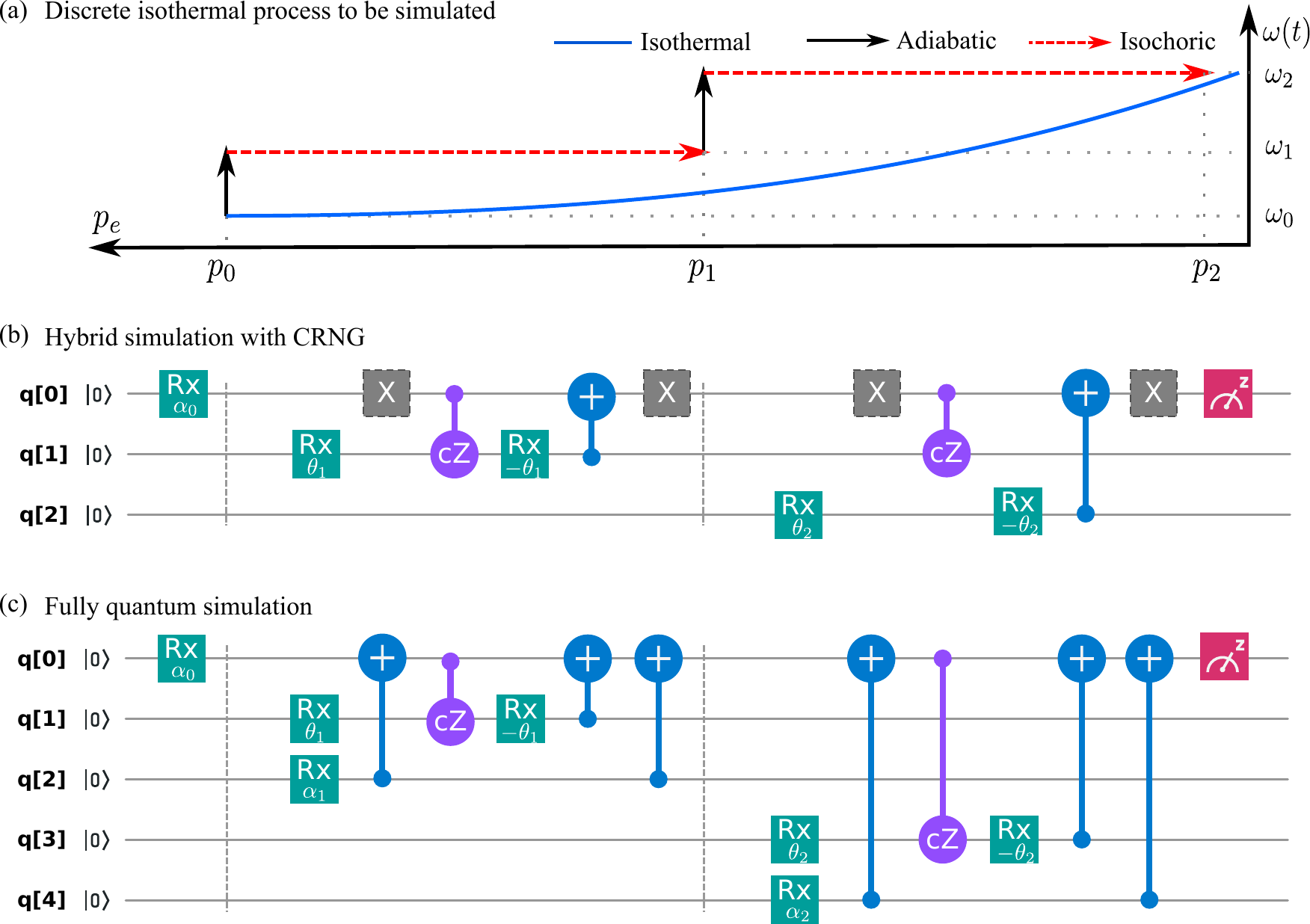}

\caption{The ibmqx2 circuit of the two-step isothermal process for the hybrid
simulation and the fully quantum simulation. (a) Excited state population-energy
($p_{e}-E$) diagram of the two-step isothermal process. (b) The circuit
for the hybrid simulation on ibmqx2. In each elementary process, the
X gate is (or not) implemented for the sub-channel selected as the
amplitude pumping (damping) channel according to the classical random
number. Each elementary process requires one another ancillary qubit.
(c) The circuit for the fully quantum simulation on ibmqx2. Each elementary
requires two ancillary qubits. \label{fig:The-example-circuit}}
\end{figure*}

With the simulation of the elementary process, we can put together
the discrete-step approach to the isothermal process with the quantum
circuit. In Fig. \ref{fig:The-example-circuit}, the circuit for the
two-step isothermal process is shown as an example. Figure \ref{fig:The-example-circuit}(a)
shows the excited state population $p_{e}(t)$ with the modulated
energy level spacing $\omega(t)$ in a two-step isothermal process.
The energy is increased from $\omega_{0}$ to $\omega_{2}$ in two
discrete steps, while the excited state population decreases from
$p_{0}$ to $p_{2}$.

Figure \ref{fig:The-example-circuit}(b) shows the quantum circuit
for the hybrid simulation on ibmqx2. With the five qubits, it is feasible
to simulate a four-step isothermal process on ibmqx2. Due to the limitation
of the qubit number, the initial state is prepared with coherence.
As stated in the description of the isochoric process, such coherence
does not affect the final state population so long as the initial
state is prepared with the population equal to that of the thermal
state. With an another ancillary qubit, a true thermal state without
coherence can be initially prepared.

In the hybrid simulation, the sub-channel $\mathcal{O}_{j}^{[l]}$
of each elementary process can be selected as either the amplitude
damping $\mathscr{E}_{\downarrow}^{(j)}$ or the pumping one $\mathscr{E}_{\uparrow}^{(j)}$
. For a $N$-step isothermal process, there are $2^{N}$ selections
of the sub-channels $\{\mathcal{O}_{1}^{[l]},\mathcal{O}_{2}^{[l]},...,\mathcal{O}_{j}^{[l]},...\mathcal{O}_{N}^{[l]}\}$
for the whole process. The circuit of every selection with $N=2,3$
and $4$ is implemented on ibmqx2. For each selection, the excited
state population $p_{e}^{[l]}(t_{j})$ at each step is obtained by
repeatedly implementing the circuit and measuring the state of the
system qubit. The work in each selection, namely the microscopic work
is given by 
\begin{equation}
W^{[l]}=\sum_{j=1}^{N}(\omega_{j}-\omega_{j-1})p_{e}^{[l]}(t_{j-1}).\label{eq:microscopicwork}
\end{equation}
The performed work $\bar{W}$ of the whole process is the average
of the microscopic work $W^{[l]}.$

Figure \ref{fig:The-example-circuit}(c) shows the fully quantum simulation
realized on ibmqx2. With the five qubits, at most two-step isothermal
process can be realized with the fully quantum simulation, since the
qubit resetting process is unavailable on ibmqx2. In fully quantum
simulation, the fixed circuit is implemented repetitively, and the
excited state population $p_{e}(t_{j})$ is obtained by measuring
the state of the system qubit. The performed work for the simulated
system is given by

\begin{equation}
\overline{W}=\sum_{j=1}^{N}(\omega_{j}-\omega_{j-1})p_{e}(t_{j-1}).\label{eq:average work}
\end{equation}

Since ibmqx2 does not allow the user to reset the state of the qubit,
each elementary process needs new additional ancillary qubit(s). If
it is available to reset the ancillary qubit, two (three) qubits are
enough to complete the simulation with the hybrid simulation (fully
quantum simulation) by resetting the ancillary qubit(s) at the end
of each isochoric process. This control scheme is realized in Ref.
\citep{Hu2018} to simulate repetitive quantum channels on a single
qubit.

\section{Testing $\mathcal{C}/\tau$ scaling of extra work\label{sec:Testing--}}

One possible application of the thermodynamic simulation is to test
the $\mathcal{C}/\tau$ scaling of the extra work. In equilibrium
thermodynamics, the work done for an ideal isothermal process is equal
to the change of the free energy $\Delta F$ \citep{book:18204}.
The ideal isothermal process requires infinite control time to ensure
the equilibrium at every moment and avoid irreversiblity. For a real
isothermal process, the irreversibility arises and the extra work
is needed to complete the process in finite time. For a fixed control
scheme, it is proved that the extra work decreases inverse proportional
to the control time at the long time limit \citep{Salamon1983}. Such
$\mathcal{C}/\tau$ scaling has been verified for the compression
of dry air in experiment \citep{Ma2019}.

The thermodynamic simulation with the quantum circuit provides an
experimental proposal to study quantum thermodynamics. We demonstrate
the scaling behavior of the extra work in finite-time isothermal process
can be observed with the current experimental proposal. The parameters
of the simulated system can be arbitrarily chosen. Here, the parameters
of the simulated two-level systems are chosen as $\gamma_{0}=1$ and
$\beta=1$ for convenience. The energy level spacing is modulated
from $\omega_{0}=1$ to $\omega_{N}=2$ in $N$ steps of elementary
processes.

\begin{figure}
\includegraphics[width=7.5cm]{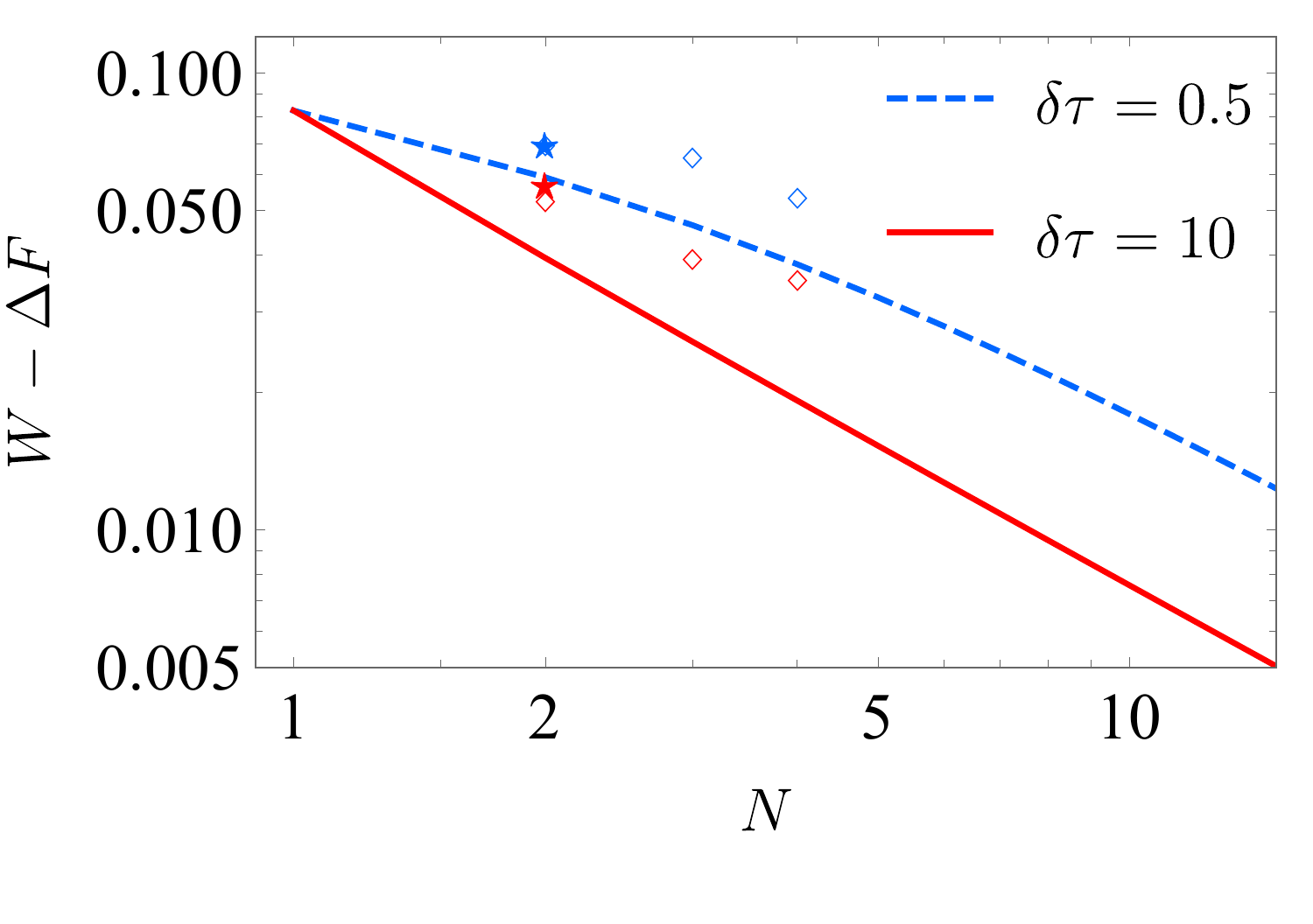}

\caption{$\mathcal{C}/N$ scaling of the extra work for the discrete isothermal
process. The time consuming of each isochoric process is set as $\delta\tau=0.5$
(blue dashed curve) or $10$ (red solid curve). The ibmqx2 results
for $N=2,3$ and $4$ are plotted, with the empty squares representing
the result of the hybrid simulations and the pentagrams representing
that of the fully quantum simulation. \label{fig:(a)-the-work}}
\end{figure}

In Fig. \ref{fig:(a)-the-work}, the $\mathcal{C}/N$ scaling of the
extra work is shown with the simulation results using the IBM quantum
computer with different time consuming $\delta\tau=0.5$ (blue dashed
curve) and $10$ (red solid curve). For large step number $N$, it
is observed that the extra work is inverse proportional to the step
number as $\overline{W}-\Delta F\propto\mathcal{C}/N$. The free energy
difference of the final and the initial state, namely the performed
work in the ideal isothermal process is

\begin{align}
\Delta F & =\omega_{N}-\omega_{0}-k_{\mathrm{B}}T\ln\frac{1+e^{\beta\omega_{N}}}{1+e^{\beta\omega_{0}}},
\end{align}
with the explicit value $\Delta F=0.186$. Since the total time consuming
is $\tau=N\delta\tau$, the $\mathcal{C}/N$ scaling implies the $\mathcal{C}/\tau$
scaling of the extra work for the finite-time discrete isothermal
process. The discrete isothermal process is simulated on ibmqx2 for
$N=2,3$ and $4$ with the hybrid simulation and $N=2$ with the fully
quantum simulation, shown as the empty squares and pentagrams, respectively.

\begin{figure}
\includegraphics[width=9cm]{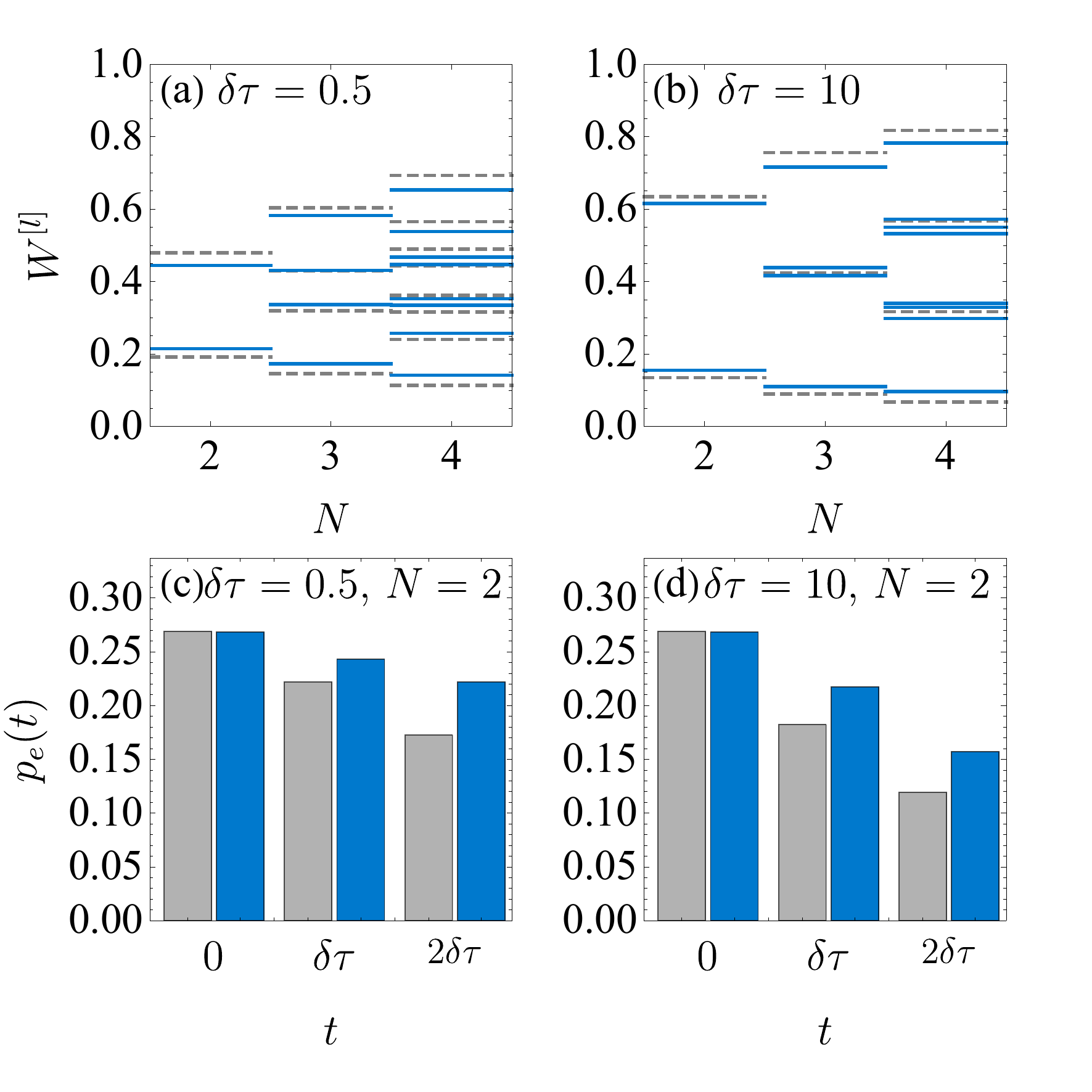}

\caption{The ibmqx2 simulation results of both the hybrid simulation and the
fully quantum simulation. The time consuming of the isochoric process
is chosen as $\delta\tau=0.5$ and $10$. (a) and (b) show the microscopic
work in the hybrid simulation of few-step isothermal processes with
$N=2,\,3$ and $4$. The ibmqx2 simulation result (blue solid line)
is compared with the numerical result (gray dashed line). (c) and
(d) give the excited state population $p_{e}(t)$ at each step in
the fully quantum simulation of the two-step isothermal process. The
ibmqx2 results (blue bar) are compared to the numerical results (gray
bar). \label{fig:The-ibmqx2-simulation}}
\end{figure}

Figure \ref{fig:The-ibmqx2-simulation} gives a detailed analysis
for the simulation results by ibmqx2. In Fig. \ref{fig:The-ibmqx2-simulation}
(a) and (b), the work distribution of the hybrid simulation results
(blue solid line) is compared to the exact numerical results (gray
dashed line), with the time consuming $\delta\tau=0.5$ in (a) and
$\delta\tau=10$ in (b). For the hybrid simulation on ibmqx2, the
maximum step number is $N=4$ with the five qubits. To mimic the random
selection of the sub-channel, we simulate every possible selection
of the sub-channels in the isochoric processes on ibmqx2 and measure
the state population of the system qubit. For each selection, the
corresponding circuit is implemented on ibmqx2 for 8192 shots. The
average work is obtained by summing the work in each selection with
the corresponding probability $p_{\{K_{j}\}}=\prod_{j}p_{K_{j}}^{(j)}$
($K_{j}=\uparrow\:\mathrm{or}\:\downarrow$). If the random selections
of the sub-channels are possible, $p_{\{K_{j}\}}$ should be determined
by the CRNG. Yet, here it is not implemented in the experiment but
calculated with $p_{K_{j}}^{(j)}$, since the random selection of
the two sub-channels is not feasible on ibmqx2.

Figure \ref{fig:The-ibmqx2-simulation} (c) and (d) shows the excited
state population of the system qubit for the fully quantum simulation
of two-step isothermal process on ibmqx2. The time consuming of each
isochoric process is set as $\delta\tau=0.5$ in (c) and $\delta\tau=10$
in (d). By measuring the state of the system qubit at each step, the
excited state population $p_{e}(t_{j})$ with $t_{j}=0,\:\delta\tau$
and $2\delta\tau$ is obtained by implementing 40960 shots in of the
corresponding circuits. Compared to that of the numerical result (gray
bar), the excited state population $p_{e}(t)$ of the ibmqx2 result
(blue bar) is larger due to the noise in the quantum computer. At
the end $t=2\delta\tau$ of the process, the most quantum gates are
used, and the absolute error reaches about $0.05$.

The performed work of the ibmqx2 results for both the hybrid simulation
and the fully quantum simulation are listed in Table. \ref{tab:The-performed-work}.
In Fig. \ref{fig:(a)-the-work}, the extra work in the ibmqx2 results
is larger than that of the numerical result since the error of the
excited state population is accumulated with the longer circuit. The
error might mainly comes from the two-qubit gates, since the error
rate of the two-qubit gates (from 1.2e-2 to 3.1e-2) is much greater
than that of the single gate (from 3.0e-4 to 1.3e-3) \citep{ibm}.
The computing accuracy can be improved by using either quantum error
correction or quantum mitigation \citep{Song2019}. Limited to the
precision of operation on ibmqx2, the results deviate from the theoretical
expectation values.

\begin{table}
\begin{tabular}{>{\raggedright}p{0.26\linewidth}>{\raggedright}m{0.06\linewidth}>{\raggedright}m{0.14\linewidth}>{\raggedright}m{0.14\linewidth}>{\raggedright}m{0.14\linewidth}>{\raggedright}m{0.14\linewidth}>{\raggedright\arraybackslash}p{0.26\linewidth}>{\raggedright\arraybackslash}p{0.26\linewidth}>{\raggedright\arraybackslash}p{0.26\linewidth}}
\cmidrule{1-6} \cmidrule{2-6} \cmidrule{3-6} \cmidrule{4-6} \cmidrule{5-6} \cmidrule{6-6} 
\multirow{1}{0.26\linewidth}{} & \multirow{2}{0.06\linewidth}{$N$} & \multicolumn{2}{l}{$\delta\tau=0.5$} & \multicolumn{2}{l}{$\delta\tau=10$} & \multicolumn{3}{c}{}\tabularnewline
\cmidrule{3-6} \cmidrule{4-6} \cmidrule{5-6} \cmidrule{6-6} 
 &  & $\overline{W}_{\mathrm{ibmqx2}}$ & $\overline{W}_{\mathrm{exact}}$ & $\overline{W}_{\mathrm{ibmqx2}}$ & $\overline{W}_{\mathrm{exact}}$ & \multicolumn{3}{c}{}\tabularnewline
\cmidrule{1-6} \cmidrule{2-6} \cmidrule{3-6} \cmidrule{4-6} \cmidrule{5-6} \cmidrule{6-6} 
\multirow{3}{0.26\linewidth}{Hybrid simulation} & 2 & 0.257 & 0.245 & 0.239 & 0.226 & \multicolumn{3}{c}{}\tabularnewline
 & 3 & 0.252 & 0.232 & 0.226 & 0.212 & \multicolumn{3}{c}{}\tabularnewline
 & 4 & 0.240 & 0.224 & 0.222 & 0.206 & \multicolumn{3}{c}{}\tabularnewline
\cmidrule{1-6} \cmidrule{2-6} \cmidrule{3-6} \cmidrule{4-6} \cmidrule{5-6} \cmidrule{6-6} 
Fully quantum simulation & 2 & 0.256 & 0.245 & 0.243 & 0.226 & \multicolumn{3}{c}{}\tabularnewline
\cmidrule{1-6} \cmidrule{2-6} \cmidrule{3-6} \cmidrule{4-6} \cmidrule{5-6} \cmidrule{6-6} 
\end{tabular}\caption{The performed work of the ibmqx2 results and the numerical results.\label{tab:The-performed-work}}
\end{table}

In the simulation, the Hamiltonian is commutative at different steps
$[H(t_{j}),H(t_{l})]=0$, and the adiabatic process is considered
as a quench with zero time consuming $\delta\tau_{\mathrm{adi}}=0$.
The current proposal can be generalized to the discrete isothermal
process with finite-time adiabatic processes. The finite-time effect
of the adiabatic process will also lead to extra cost of the performed
work for the case with the non-commutative Hamiltonian at different
time \citep{Chen2019}. For a generic adiabatic process, the unitary
evolution of the two-level system should be simulated with the single-qubit
gates on the system qubit.

With the current available number of qubits, we only show limited
data points on the scale cureve in Fig. \ref{fig:(a)-the-work}. Additional
effort can be added to test the optimal control scheme\citep{Ma2018}.
For the fixed control time $\tau$, the control scheme can be optimized
to reach the minimum extra work. The lower bound of the extra work
is related to the concept of the thermodynamic length \citep{Salamon1983,Deffner2013,Crooks2007,Scandi2018}.
Such concept endows a Riemann metric on the space of the tuning parameters.
The current experimental proposal might also be utilized to measure
the thermodynamic length of the isothermal process for the two-level
system.

\section{Conclusion\label{sec:Conclusion}}

We show a proposal to simulate the finite-time isothermal process
of two-level system using the superconducting quantum computer. Two
methods, the hybrid simulation and the fully quantum simulation, are
given to realize the generalized amplitude damping channel. Assisted
by the classical random number generator or the quantum superposition,
the hybrid simulation or the fully quantum simulation can simulate
a $N$-step isothermal process with $N+1$ or $2N+1$ qubits, respectively.

We have used the quantum computer of IBM (ibmqx2) to demonstrate the
simulation of the discrete isothermal process. The discrete isothermal
process has been realized for four steps with the hybrid simulation
and two steps with the fully quantum simulation. If more steps of
elementary processes can be realized in the experiment, the $\mathcal{C}/\tau$
scaling of the extra work can be observed using the thermodynamic
simulation on the universal quantum computer.
\begin{acknowledgments}
We thank Luyan Sun for helpful discussions at the initial stage of
the current work. This work is supported by the NSFC (Grants No. 11534002,
No. 11875049 and No. 11875050), the NSAF (Grant No. U1930403 and No.
U1930402), and the National Basic Research Program of China (Grants
No. 2016YFA0301201). HD and YL also thank The Recruitment Program
of Global Youth Experts of China.
\end{acknowledgments}

\bibliographystyle{apsrev4-1}
\bibliography{simulation,main}

\begin{thebibliography}{38}%
\makeatletter
\providecommand \@ifxundefined [1]{%
 \@ifx{#1\undefined}
}%
\providecommand \@ifnum [1]{%
 \ifnum #1\expandafter \@firstoftwo
 \else \expandafter \@secondoftwo
 \fi
}%
\providecommand \@ifx [1]{%
 \ifx #1\expandafter \@firstoftwo
 \else \expandafter \@secondoftwo
 \fi
}%
\providecommand \natexlab [1]{#1}%
\providecommand \enquote  [1]{``#1''}%
\providecommand \bibnamefont  [1]{#1}%
\providecommand \bibfnamefont [1]{#1}%
\providecommand \citenamefont [1]{#1}%
\providecommand \href@noop [0]{\@secondoftwo}%
\providecommand \href [0]{\begingroup \@sanitize@url \@href}%
\providecommand \@href[1]{\@@startlink{#1}\@@href}%
\providecommand \@@href[1]{\endgroup#1\@@endlink}%
\providecommand \@sanitize@url [0]{\catcode `\\12\catcode `\$12\catcode
  `\&12\catcode `\#12\catcode `\^12\catcode `\_12\catcode `\%12\relax}%
\providecommand \@@startlink[1]{}%
\providecommand \@@endlink[0]{}%
\providecommand \url  [0]{\begingroup\@sanitize@url \@url }%
\providecommand \@url [1]{\endgroup\@href {#1}{\urlprefix }}%
\providecommand \urlprefix  [0]{URL }%
\providecommand \Eprint [0]{\href }%
\providecommand \doibase [0]{http://dx.doi.org/}%
\providecommand \selectlanguage [0]{\@gobble}%
\providecommand \bibinfo  [0]{\@secondoftwo}%
\providecommand \bibfield  [0]{\@secondoftwo}%
\providecommand \translation [1]{[#1]}%
\providecommand \BibitemOpen [0]{}%
\providecommand \bibitemStop [0]{}%
\providecommand \bibitemNoStop [0]{.\EOS\space}%
\providecommand \EOS [0]{\spacefactor3000\relax}%
\providecommand \BibitemShut  [1]{\csname bibitem#1\endcsname}%
\let\auto@bib@innerbib\@empty
%</preamble>
\bibitem [{\citenamefont {Campisi}\ \emph {et~al.}(2011)\citenamefont
  {Campisi}, \citenamefont {H\"{a}nggi},\ and\ \citenamefont
  {Talkner}}]{Campisi_2011}%
  \BibitemOpen
  \bibfield  {author} {\bibinfo {author} {\bibfnamefont {M.}~\bibnamefont
  {Campisi}}, \bibinfo {author} {\bibfnamefont {P.}~\bibnamefont {H\"{a}nggi}},
  \ and\ \bibinfo {author} {\bibfnamefont {P.}~\bibnamefont {Talkner}},\ }\href
  {\doibase 10.1103/revmodphys.83.771} {\bibfield  {journal} {\bibinfo
  {journal} {Rev. Mod. Phys.}\ }\textbf {\bibinfo {volume} {83}},\ \bibinfo
  {pages} {771} (\bibinfo {year} {2011})}\BibitemShut {NoStop}%
\bibitem [{\citenamefont {Esposito}\ \emph {et~al.}(2009)\citenamefont
  {Esposito}, \citenamefont {Harbola},\ and\ \citenamefont
  {Mukamel}}]{Esposito_2009}%
  \BibitemOpen
  \bibfield  {author} {\bibinfo {author} {\bibfnamefont {M.}~\bibnamefont
  {Esposito}}, \bibinfo {author} {\bibfnamefont {U.}~\bibnamefont {Harbola}}, \
  and\ \bibinfo {author} {\bibfnamefont {S.}~\bibnamefont {Mukamel}},\ }\href
  {\doibase 10.1103/revmodphys.81.1665} {\bibfield  {journal} {\bibinfo
  {journal} {Rev. Mod. Phys.}\ }\textbf {\bibinfo {volume} {81}},\ \bibinfo
  {pages} {1665} (\bibinfo {year} {2009})}\BibitemShut {NoStop}%
\bibitem [{\citenamefont {Maruyama}\ \emph {et~al.}(2009)\citenamefont
  {Maruyama}, \citenamefont {Nori},\ and\ \citenamefont
  {Vedral}}]{Maruyama2009}%
  \BibitemOpen
  \bibfield  {author} {\bibinfo {author} {\bibfnamefont {K.}~\bibnamefont
  {Maruyama}}, \bibinfo {author} {\bibfnamefont {F.}~\bibnamefont {Nori}}, \
  and\ \bibinfo {author} {\bibfnamefont {V.}~\bibnamefont {Vedral}},\ }\href
  {\doibase 10.1103/RevModPhys.81.1} {\bibfield  {journal} {\bibinfo  {journal}
  {Rev. Mod. Phys.}\ }\textbf {\bibinfo {volume} {81}},\ \bibinfo {pages} {1}
  (\bibinfo {year} {2009})}\BibitemShut {NoStop}%
\bibitem [{\citenamefont {Linden}\ \emph {et~al.}(2010)\citenamefont {Linden},
  \citenamefont {Popescu},\ and\ \citenamefont
  {Skrzypczyk}}]{Linden2010PhysRevLett105_130401}%
  \BibitemOpen
  \bibfield  {author} {\bibinfo {author} {\bibfnamefont {N.}~\bibnamefont
  {Linden}}, \bibinfo {author} {\bibfnamefont {S.}~\bibnamefont {Popescu}}, \
  and\ \bibinfo {author} {\bibfnamefont {P.}~\bibnamefont {Skrzypczyk}},\
  }\href {\doibase 10.1103/physrevlett.105.130401} {\bibfield  {journal}
  {\bibinfo  {journal} {Phys. Rev. Lett.}\ }\textbf {\bibinfo {volume} {105}},\
  \bibinfo {pages} {130401} (\bibinfo {year} {2010})}\BibitemShut {NoStop}%
\bibitem [{\citenamefont {Strasberg}\ \emph {et~al.}(2017)\citenamefont
  {Strasberg}, \citenamefont {Schaller}, \citenamefont {Brandes},\ and\
  \citenamefont {Esposito}}]{Strasberg_2017}%
  \BibitemOpen
  \bibfield  {author} {\bibinfo {author} {\bibfnamefont {P.}~\bibnamefont
  {Strasberg}}, \bibinfo {author} {\bibfnamefont {G.}~\bibnamefont {Schaller}},
  \bibinfo {author} {\bibfnamefont {T.}~\bibnamefont {Brandes}}, \ and\
  \bibinfo {author} {\bibfnamefont {M.}~\bibnamefont {Esposito}},\ }\href
  {\doibase 10.1103/physrevx.7.021003} {\bibfield  {journal} {\bibinfo
  {journal} {Phys. Rev. X}\ }\textbf {\bibinfo {volume} {7}},\ \bibinfo {pages}
  {021003} (\bibinfo {year} {2017})}\BibitemShut {NoStop}%
\bibitem [{\citenamefont {Vinjanampathy}\ and\ \citenamefont
  {Anders}(2016)}]{Vinjanampathy_2016}%
  \BibitemOpen
  \bibfield  {author} {\bibinfo {author} {\bibfnamefont {S.}~\bibnamefont
  {Vinjanampathy}}\ and\ \bibinfo {author} {\bibfnamefont {J.}~\bibnamefont
  {Anders}},\ }\href {\doibase 10.1080/00107514.2016.1201896} {\bibfield
  {journal} {\bibinfo  {journal} {Contemp. Phys.}\ }\textbf {\bibinfo {volume}
  {57}},\ \bibinfo {pages} {545} (\bibinfo {year} {2016})}\BibitemShut
  {NoStop}%
\bibitem [{\citenamefont {Feynman}(1982)}]{Feynman1982}%
  \BibitemOpen
  \bibfield  {author} {\bibinfo {author} {\bibfnamefont {R.~P.}\ \bibnamefont
  {Feynman}},\ }\href {\doibase 10.1007/bf02650179} {\bibfield  {journal}
  {\bibinfo  {journal} {Int. J. Theor. Phys.}\ }\textbf {\bibinfo {volume}
  {21}},\ \bibinfo {pages} {467} (\bibinfo {year} {1982})}\BibitemShut
  {NoStop}%
\bibitem [{\citenamefont {Lloyd}(1996)}]{Lloyd1996}%
  \BibitemOpen
  \bibfield  {author} {\bibinfo {author} {\bibfnamefont {S.}~\bibnamefont
  {Lloyd}},\ }\href {\doibase 10.1126/science.273.5278.1073} {\bibfield
  {journal} {\bibinfo  {journal} {Science}\ }\textbf {\bibinfo {volume}
  {273}},\ \bibinfo {pages} {1073} (\bibinfo {year} {1996})}\BibitemShut
  {NoStop}%
\bibitem [{\citenamefont {Georgescu}\ \emph {et~al.}(2014)\citenamefont
  {Georgescu}, \citenamefont {Ashhab},\ and\ \citenamefont
  {Nori}}]{Georgescu2014}%
  \BibitemOpen
  \bibfield  {author} {\bibinfo {author} {\bibfnamefont {I.}~\bibnamefont
  {Georgescu}}, \bibinfo {author} {\bibfnamefont {S.}~\bibnamefont {Ashhab}}, \
  and\ \bibinfo {author} {\bibfnamefont {F.}~\bibnamefont {Nori}},\ }\href
  {\doibase 10.1103/revmodphys.86.153} {\bibfield  {journal} {\bibinfo
  {journal} {Rev. Mod. Phys.}\ }\textbf {\bibinfo {volume} {86}},\ \bibinfo
  {pages} {153} (\bibinfo {year} {2014})}\BibitemShut {NoStop}%
\bibitem [{\citenamefont {Sandholzer}\ \emph {et~al.}(2019)\citenamefont
  {Sandholzer}, \citenamefont {Murakami}, \citenamefont {G\"{o}rg},
  \citenamefont {Minguzzi}, \citenamefont {Messer}, \citenamefont {Desbuquois},
  \citenamefont {Eckstein}, \citenamefont {Werner},\ and\ \citenamefont
  {Esslinger}}]{Sandholzer2019}%
  \BibitemOpen
  \bibfield  {author} {\bibinfo {author} {\bibfnamefont {K.}~\bibnamefont
  {Sandholzer}}, \bibinfo {author} {\bibfnamefont {Y.}~\bibnamefont
  {Murakami}}, \bibinfo {author} {\bibfnamefont {F.}~\bibnamefont {G\"{o}rg}},
  \bibinfo {author} {\bibfnamefont {J.}~\bibnamefont {Minguzzi}}, \bibinfo
  {author} {\bibfnamefont {M.}~\bibnamefont {Messer}}, \bibinfo {author}
  {\bibfnamefont {R.}~\bibnamefont {Desbuquois}}, \bibinfo {author}
  {\bibfnamefont {M.}~\bibnamefont {Eckstein}}, \bibinfo {author}
  {\bibfnamefont {P.}~\bibnamefont {Werner}}, \ and\ \bibinfo {author}
  {\bibfnamefont {T.}~\bibnamefont {Esslinger}},\ }\href {\doibase
  10.1103/physrevlett.123.193602} {\bibfield  {journal} {\bibinfo  {journal}
  {Phys. Rev. Lett.}\ }\textbf {\bibinfo {volume} {123}},\ \bibinfo {pages}
  {193602} (\bibinfo {year} {2019})}\BibitemShut {NoStop}%
\bibitem [{\citenamefont {An}\ \emph {et~al.}(2014)\citenamefont {An},
  \citenamefont {Zhang}, \citenamefont {Um}, \citenamefont {Lv}, \citenamefont
  {Lu}, \citenamefont {Zhang}, \citenamefont {Yin}, \citenamefont {Quan},\ and\
  \citenamefont {Kim}}]{An2014}%
  \BibitemOpen
  \bibfield  {author} {\bibinfo {author} {\bibfnamefont {S.}~\bibnamefont
  {An}}, \bibinfo {author} {\bibfnamefont {J.-N.}\ \bibnamefont {Zhang}},
  \bibinfo {author} {\bibfnamefont {M.}~\bibnamefont {Um}}, \bibinfo {author}
  {\bibfnamefont {D.}~\bibnamefont {Lv}}, \bibinfo {author} {\bibfnamefont
  {Y.}~\bibnamefont {Lu}}, \bibinfo {author} {\bibfnamefont {J.}~\bibnamefont
  {Zhang}}, \bibinfo {author} {\bibfnamefont {Z.-Q.}\ \bibnamefont {Yin}},
  \bibinfo {author} {\bibfnamefont {H.~T.}\ \bibnamefont {Quan}}, \ and\
  \bibinfo {author} {\bibfnamefont {K.}~\bibnamefont {Kim}},\ }\href {\doibase
  10.1038/nphys3197} {\bibfield  {journal} {\bibinfo  {journal} {Nat. Phys.}\
  }\textbf {\bibinfo {volume} {11}},\ \bibinfo {pages} {193} (\bibinfo {year}
  {2014})}\BibitemShut {NoStop}%
\bibitem [{\citenamefont {Deng}\ \emph {et~al.}(2015)\citenamefont {Deng},
  \citenamefont {Diao}, \citenamefont {Yu},\ and\ \citenamefont
  {Wu}}]{Deng2015}%
  \BibitemOpen
  \bibfield  {author} {\bibinfo {author} {\bibfnamefont {S.-J.}\ \bibnamefont
  {Deng}}, \bibinfo {author} {\bibfnamefont {P.-P.}\ \bibnamefont {Diao}},
  \bibinfo {author} {\bibfnamefont {Q.-L.}\ \bibnamefont {Yu}}, \ and\ \bibinfo
  {author} {\bibfnamefont {H.-B.}\ \bibnamefont {Wu}},\ }\href {\doibase
  10.1088/0256-307x/32/5/053401} {\bibfield  {journal} {\bibinfo  {journal}
  {Chin. Phys. Lett.}\ }\textbf {\bibinfo {volume} {32}},\ \bibinfo {pages}
  {053401} (\bibinfo {year} {2015})}\BibitemShut {NoStop}%
\bibitem [{\citenamefont {Wang}\ \emph {et~al.}(2019)\citenamefont {Wang},
  \citenamefont {Zhang}, \citenamefont {Xiang}, \citenamefont {Jia},
  \citenamefont {Duan}, \citenamefont {Zong}, \citenamefont {Sun},
  \citenamefont {Dong}, \citenamefont {Wu}, \citenamefont {Yin},\ and\
  \citenamefont {Guo}}]{Wang2019}%
  \BibitemOpen
  \bibfield  {author} {\bibinfo {author} {\bibfnamefont {T.}~\bibnamefont
  {Wang}}, \bibinfo {author} {\bibfnamefont {Z.}~\bibnamefont {Zhang}},
  \bibinfo {author} {\bibfnamefont {L.}~\bibnamefont {Xiang}}, \bibinfo
  {author} {\bibfnamefont {Z.}~\bibnamefont {Jia}}, \bibinfo {author}
  {\bibfnamefont {P.}~\bibnamefont {Duan}}, \bibinfo {author} {\bibfnamefont
  {Z.}~\bibnamefont {Zong}}, \bibinfo {author} {\bibfnamefont {Z.}~\bibnamefont
  {Sun}}, \bibinfo {author} {\bibfnamefont {Z.}~\bibnamefont {Dong}}, \bibinfo
  {author} {\bibfnamefont {J.}~\bibnamefont {Wu}}, \bibinfo {author}
  {\bibfnamefont {Y.}~\bibnamefont {Yin}}, \ and\ \bibinfo {author}
  {\bibfnamefont {G.}~\bibnamefont {Guo}},\ }\href {\doibase
  10.1103/physrevapplied.11.034030} {\bibfield  {journal} {\bibinfo  {journal}
  {Phys. Rev. Appl}\ }\textbf {\bibinfo {volume} {11}},\ \bibinfo {pages}
  {034030} (\bibinfo {year} {2019})}\BibitemShut {NoStop}%
\bibitem [{\citenamefont {Alicki}(1979)}]{Alicki1979}%
  \BibitemOpen
  \bibfield  {author} {\bibinfo {author} {\bibfnamefont {R.}~\bibnamefont
  {Alicki}},\ }\href {\doibase 10.1088/0305-4470/12/5/007} {\bibfield
  {journal} {\bibinfo  {journal} {J. Phys. A: Math. Gen.}\ }\textbf {\bibinfo
  {volume} {12}},\ \bibinfo {pages} {L103} (\bibinfo {year}
  {1979})}\BibitemShut {NoStop}%
\bibitem [{\citenamefont {Quan}\ \emph {et~al.}(2007)\citenamefont {Quan},
  \citenamefont {Liu}, \citenamefont {Sun},\ and\ \citenamefont
  {Nori}}]{Quan_2007}%
  \BibitemOpen
  \bibfield  {author} {\bibinfo {author} {\bibfnamefont {H.~T.}\ \bibnamefont
  {Quan}}, \bibinfo {author} {\bibfnamefont {Y.~X.}\ \bibnamefont {Liu}},
  \bibinfo {author} {\bibfnamefont {C.~P.}\ \bibnamefont {Sun}}, \ and\
  \bibinfo {author} {\bibfnamefont {F.}~\bibnamefont {Nori}},\ }\href {\doibase
  10.1103/physreve.76.031105} {\bibfield  {journal} {\bibinfo  {journal} {Phys.
  Rev. E}\ }\textbf {\bibinfo {volume} {76}},\ \bibinfo {pages} {031105}
  (\bibinfo {year} {2007})}\BibitemShut {NoStop}%
\bibitem [{\citenamefont {Su}\ \emph {et~al.}(2018)\citenamefont {Su},
  \citenamefont {Chen}, \citenamefont {Ma}, \citenamefont {Chen},\ and\
  \citenamefont {Sun}}]{Su2018}%
  \BibitemOpen
  \bibfield  {author} {\bibinfo {author} {\bibfnamefont {S.}~\bibnamefont
  {Su}}, \bibinfo {author} {\bibfnamefont {J.}~\bibnamefont {Chen}}, \bibinfo
  {author} {\bibfnamefont {Y.}~\bibnamefont {Ma}}, \bibinfo {author}
  {\bibfnamefont {J.}~\bibnamefont {Chen}}, \ and\ \bibinfo {author}
  {\bibfnamefont {C.}~\bibnamefont {Sun}},\ }\href {\doibase
  10.1088/1674-1056/27/6/060502} {\bibfield  {journal} {\bibinfo  {journal}
  {Chin. Phys. B}\ }\textbf {\bibinfo {volume} {27}},\ \bibinfo {pages}
  {060502} (\bibinfo {year} {2018})}\BibitemShut {NoStop}%
\bibitem [{\citenamefont {Blatt}\ and\ \citenamefont {Roos}(2012)}]{Blatt2012}%
  \BibitemOpen
  \bibfield  {author} {\bibinfo {author} {\bibfnamefont {R.}~\bibnamefont
  {Blatt}}\ and\ \bibinfo {author} {\bibfnamefont {C.~F.}\ \bibnamefont
  {Roos}},\ }\href {\doibase 10.1038/nphys2252} {\bibfield  {journal} {\bibinfo
   {journal} {Nat. Phys.}\ }\textbf {\bibinfo {volume} {8}},\ \bibinfo {pages}
  {277} (\bibinfo {year} {2012})}\BibitemShut {NoStop}%
\bibitem [{\citenamefont {Talkner}\ and\ \citenamefont
  {H\"{a}nggi}(2016)}]{Talkner2016}%
  \BibitemOpen
  \bibfield  {author} {\bibinfo {author} {\bibfnamefont {P.}~\bibnamefont
  {Talkner}}\ and\ \bibinfo {author} {\bibfnamefont {P.}~\bibnamefont
  {H\"{a}nggi}},\ }\href {\doibase 10.1103/physreve.93.022131} {\bibfield
  {journal} {\bibinfo  {journal} {Phys. Rev. E}\ }\textbf {\bibinfo {volume}
  {93}},\ \bibinfo {pages} {022131} (\bibinfo {year} {2016})}\BibitemShut
  {NoStop}%
\bibitem [{ibm()}]{ibm}%
  \BibitemOpen
  \href {http://research.ibm.com/ibm-q/} {\enquote {\bibinfo {title} {{IBM}
  {Q}uantum {E}xperience},}\ }\BibitemShut {NoStop}%
\bibitem [{\citenamefont {Ruskai}\ \emph {et~al.}(2002)\citenamefont {Ruskai},
  \citenamefont {Szarek},\ and\ \citenamefont {Werner}}]{Ruskai2002}%
  \BibitemOpen
  \bibfield  {author} {\bibinfo {author} {\bibfnamefont {M.~B.}\ \bibnamefont
  {Ruskai}}, \bibinfo {author} {\bibfnamefont {S.}~\bibnamefont {Szarek}}, \
  and\ \bibinfo {author} {\bibfnamefont {E.}~\bibnamefont {Werner}},\ }\href
  {\doibase 10.1016/s0024-3795(01)00547-x} {\bibfield  {journal} {\bibinfo
  {journal} {Linear Algebra Appl}\ }\textbf {\bibinfo {volume} {347}},\
  \bibinfo {pages} {159} (\bibinfo {year} {2002})}\BibitemShut {NoStop}%
\bibitem [{\citenamefont {Nielsen}\ and\ \citenamefont
  {Chuang}(2009)}]{Nielsen2009}%
  \BibitemOpen
  \bibfield  {author} {\bibinfo {author} {\bibfnamefont {M.~A.}\ \bibnamefont
  {Nielsen}}\ and\ \bibinfo {author} {\bibfnamefont {I.~L.}\ \bibnamefont
  {Chuang}},\ }\href {\doibase 10.1017/cbo9780511976667} {\emph {\bibinfo
  {title} {Quantum Computation and Quantum Information}}}\ (\bibinfo
  {publisher} {Cambridge University Press},\ \bibinfo {year}
  {2009})\BibitemShut {NoStop}%
\bibitem [{\citenamefont {Fisher}\ \emph {et~al.}(2012)\citenamefont {Fisher},
  \citenamefont {Prevedel}, \citenamefont {Kaltenbaek},\ and\ \citenamefont
  {Resch}}]{Fisher2012}%
  \BibitemOpen
  \bibfield  {author} {\bibinfo {author} {\bibfnamefont {K.~A.~G.}\
  \bibnamefont {Fisher}}, \bibinfo {author} {\bibfnamefont {R.}~\bibnamefont
  {Prevedel}}, \bibinfo {author} {\bibfnamefont {R.}~\bibnamefont
  {Kaltenbaek}}, \ and\ \bibinfo {author} {\bibfnamefont {K.~J.}\ \bibnamefont
  {Resch}},\ }\href {\doibase 10.1088/1367-2630/14/3/033016} {\bibfield
  {journal} {\bibinfo  {journal} {New J. Phys.}\ }\textbf {\bibinfo {volume}
  {14}},\ \bibinfo {pages} {033016} (\bibinfo {year} {2012})}\BibitemShut
  {NoStop}%
\bibitem [{\citenamefont {Wang}\ \emph {et~al.}(2013)\citenamefont {Wang},
  \citenamefont {Berry}, \citenamefont {de~Oliveira},\ and\ \citenamefont
  {Sanders}}]{Wang2013}%
  \BibitemOpen
  \bibfield  {author} {\bibinfo {author} {\bibfnamefont {D.-S.}\ \bibnamefont
  {Wang}}, \bibinfo {author} {\bibfnamefont {D.~W.}\ \bibnamefont {Berry}},
  \bibinfo {author} {\bibfnamefont {M.~C.}\ \bibnamefont {de~Oliveira}}, \ and\
  \bibinfo {author} {\bibfnamefont {B.~C.}\ \bibnamefont {Sanders}},\ }\href
  {\doibase 10.1103/physrevlett.111.130504} {\bibfield  {journal} {\bibinfo
  {journal} {Phys. Rev. Lett.}\ }\textbf {\bibinfo {volume} {111}},\ \bibinfo
  {pages} {130504} (\bibinfo {year} {2013})}\BibitemShut {NoStop}%
\bibitem [{\citenamefont {Lu}\ \emph {et~al.}(2017)\citenamefont {Lu},
  \citenamefont {Liu}, \citenamefont {Wang}, \citenamefont {Chen},
  \citenamefont {Li}, \citenamefont {Yao}, \citenamefont {Li}, \citenamefont
  {Liu}, \citenamefont {Peng}, \citenamefont {Sanders}, \citenamefont {Chen},\
  and\ \citenamefont {Pan}}]{Lu2017}%
  \BibitemOpen
  \bibfield  {author} {\bibinfo {author} {\bibfnamefont {H.}~\bibnamefont
  {Lu}}, \bibinfo {author} {\bibfnamefont {C.}~\bibnamefont {Liu}}, \bibinfo
  {author} {\bibfnamefont {D.-S.}\ \bibnamefont {Wang}}, \bibinfo {author}
  {\bibfnamefont {L.-K.}\ \bibnamefont {Chen}}, \bibinfo {author}
  {\bibfnamefont {Z.-D.}\ \bibnamefont {Li}}, \bibinfo {author} {\bibfnamefont
  {X.-C.}\ \bibnamefont {Yao}}, \bibinfo {author} {\bibfnamefont
  {L.}~\bibnamefont {Li}}, \bibinfo {author} {\bibfnamefont {N.-L.}\
  \bibnamefont {Liu}}, \bibinfo {author} {\bibfnamefont {C.-Z.}\ \bibnamefont
  {Peng}}, \bibinfo {author} {\bibfnamefont {B.~C.}\ \bibnamefont {Sanders}},
  \bibinfo {author} {\bibfnamefont {Y.-A.}\ \bibnamefont {Chen}}, \ and\
  \bibinfo {author} {\bibfnamefont {J.-W.}\ \bibnamefont {Pan}},\ }\href
  {\doibase 10.1103/physreva.95.042310} {\bibfield  {journal} {\bibinfo
  {journal} {Phys. Rev. A}\ }\textbf {\bibinfo {volume} {95}},\ \bibinfo
  {pages} {042310} (\bibinfo {year} {2017})}\BibitemShut {NoStop}%
\bibitem [{\citenamefont {Hu}\ \emph {et~al.}(2018)\citenamefont {Hu},
  \citenamefont {Mu}, \citenamefont {Cai}, \citenamefont {Ma}, \citenamefont
  {Xu}, \citenamefont {Wang}, \citenamefont {Song}, \citenamefont {Zou},\ and\
  \citenamefont {Sun}}]{Hu2018}%
  \BibitemOpen
  \bibfield  {author} {\bibinfo {author} {\bibfnamefont {L.}~\bibnamefont
  {Hu}}, \bibinfo {author} {\bibfnamefont {X.}~\bibnamefont {Mu}}, \bibinfo
  {author} {\bibfnamefont {W.}~\bibnamefont {Cai}}, \bibinfo {author}
  {\bibfnamefont {Y.}~\bibnamefont {Ma}}, \bibinfo {author} {\bibfnamefont
  {Y.}~\bibnamefont {Xu}}, \bibinfo {author} {\bibfnamefont {H.}~\bibnamefont
  {Wang}}, \bibinfo {author} {\bibfnamefont {Y.}~\bibnamefont {Song}}, \bibinfo
  {author} {\bibfnamefont {C.-L.}\ \bibnamefont {Zou}}, \ and\ \bibinfo
  {author} {\bibfnamefont {L.}~\bibnamefont {Sun}},\ }\href {\doibase
  10.1016/j.scib.2018.11.010} {\bibfield  {journal} {\bibinfo  {journal} {Sci.
  Bull.}\ }\textbf {\bibinfo {volume} {63}},\ \bibinfo {pages} {1551} (\bibinfo
  {year} {2018})}\BibitemShut {NoStop}%
\bibitem [{\citenamefont {Wang}\ \emph {et~al.}(2011)\citenamefont {Wang},
  \citenamefont {Ashhab},\ and\ \citenamefont {Nori}}]{Wang2011}%
  \BibitemOpen
  \bibfield  {author} {\bibinfo {author} {\bibfnamefont {H.}~\bibnamefont
  {Wang}}, \bibinfo {author} {\bibfnamefont {S.}~\bibnamefont {Ashhab}}, \ and\
  \bibinfo {author} {\bibfnamefont {F.}~\bibnamefont {Nori}},\ }\href {\doibase
  10.1103/physreva.83.062317} {\bibfield  {journal} {\bibinfo  {journal} {Phys.
  Rev. A}\ }\textbf {\bibinfo {volume} {83}},\ \bibinfo {pages} {062317}
  (\bibinfo {year} {2011})}\BibitemShut {NoStop}%
\bibitem [{\citenamefont {Uzdin}\ and\ \citenamefont {Katz}()}]{Uzdin2019}%
  \BibitemOpen
  \bibfield  {author} {\bibinfo {author} {\bibfnamefont {R.}~\bibnamefont
  {Uzdin}}\ and\ \bibinfo {author} {\bibfnamefont {N.}~\bibnamefont {Katz}},\
  }\href@noop {} {\ }\Eprint {http://arxiv.org/abs/1908.08968v1} {1908.08968v1}
  \BibitemShut {NoStop}%
\bibitem [{\citenamefont {Su}\ and\ \citenamefont {Li}(2020)}]{Su2020}%
  \BibitemOpen
  \bibfield  {author} {\bibinfo {author} {\bibfnamefont {H.-Y.}\ \bibnamefont
  {Su}}\ and\ \bibinfo {author} {\bibfnamefont {Y.}~\bibnamefont {Li}},\ }\href
  {\doibase 10.1103/physreva.101.012328} {\bibfield  {journal} {\bibinfo
  {journal} {Phys. Rev. A}\ }\textbf {\bibinfo {volume} {101}},\ \bibinfo
  {pages} {012328} (\bibinfo {year} {2020})}\BibitemShut {NoStop}%
\bibitem [{\citenamefont {Quan}\ \emph {et~al.}(2008)\citenamefont {Quan},
  \citenamefont {Yang},\ and\ \citenamefont {Sun}}]{Quan_2008}%
  \BibitemOpen
  \bibfield  {author} {\bibinfo {author} {\bibfnamefont {H.~T.}\ \bibnamefont
  {Quan}}, \bibinfo {author} {\bibfnamefont {S.}~\bibnamefont {Yang}}, \ and\
  \bibinfo {author} {\bibfnamefont {C.~P.}\ \bibnamefont {Sun}},\ }\href
  {\doibase 10.1103/physreve.78.021116} {\bibfield  {journal} {\bibinfo
  {journal} {Phys. Rev. E}\ }\textbf {\bibinfo {volume} {78}},\ \bibinfo
  {pages} {021116} (\bibinfo {year} {2008})}\BibitemShut {NoStop}%
\bibitem [{\citenamefont {Ma}\ \emph {et~al.}(2018)\citenamefont {Ma},
  \citenamefont {Xu}, \citenamefont {Dong},\ and\ \citenamefont
  {Sun}}]{Ma2018}%
  \BibitemOpen
  \bibfield  {author} {\bibinfo {author} {\bibfnamefont {Y.-H.}\ \bibnamefont
  {Ma}}, \bibinfo {author} {\bibfnamefont {D.}~\bibnamefont {Xu}}, \bibinfo
  {author} {\bibfnamefont {H.}~\bibnamefont {Dong}}, \ and\ \bibinfo {author}
  {\bibfnamefont {C.-P.}\ \bibnamefont {Sun}},\ }\href {\doibase
  10.1103/physreve.98.022133} {\bibfield  {journal} {\bibinfo  {journal} {Phys.
  Rev. E}\ }\textbf {\bibinfo {volume} {98}},\ \bibinfo {pages} {022133}
  (\bibinfo {year} {2018})}\BibitemShut {NoStop}%
\bibitem [{\citenamefont {Callen}(1985)}]{book:18204}%
  \BibitemOpen
  \bibfield  {author} {\bibinfo {author} {\bibfnamefont {H.~B.}\ \bibnamefont
  {Callen}},\ }\href
  {http://gen.lib.rus.ec/book/index.php?md5=06F94005414F92297CEC04EF5BF987CD}
  {\emph {\bibinfo {title} {Thermodynamics And An Introduction To
  Thermostatistics}}},\ \bibinfo {edition} {2nd}\ ed.\ (\bibinfo  {publisher}
  {Wiley},\ \bibinfo {year} {1985})\BibitemShut {NoStop}%
\bibitem [{\citenamefont {Salamon}\ and\ \citenamefont
  {Berry}(1983)}]{Salamon1983}%
  \BibitemOpen
  \bibfield  {author} {\bibinfo {author} {\bibfnamefont {P.}~\bibnamefont
  {Salamon}}\ and\ \bibinfo {author} {\bibfnamefont {R.~S.}\ \bibnamefont
  {Berry}},\ }\href {\doibase 10.1103/physrevlett.51.1127} {\bibfield
  {journal} {\bibinfo  {journal} {Phys. Rev. Lett.}\ }\textbf {\bibinfo
  {volume} {51}},\ \bibinfo {pages} {1127} (\bibinfo {year}
  {1983})}\BibitemShut {NoStop}%
\bibitem [{\citenamefont {Ma}\ \emph {et~al.}()\citenamefont {Ma},
  \citenamefont {Zhai}, \citenamefont {Sun},\ and\ \citenamefont
  {Dong}}]{Ma2019}%
  \BibitemOpen
  \bibfield  {author} {\bibinfo {author} {\bibfnamefont {Y.-H.}\ \bibnamefont
  {Ma}}, \bibinfo {author} {\bibfnamefont {R.-X.}\ \bibnamefont {Zhai}},
  \bibinfo {author} {\bibfnamefont {C.-P.}\ \bibnamefont {Sun}}, \ and\
  \bibinfo {author} {\bibfnamefont {H.}~\bibnamefont {Dong}},\ }\href@noop {}
  {\ }\Eprint {http://arxiv.org/abs/1910.13434v3} {1910.13434v3} \BibitemShut
  {NoStop}%
\bibitem [{\citenamefont {Song}\ \emph {et~al.}(2019)\citenamefont {Song},
  \citenamefont {Cui}, \citenamefont {Wang}, \citenamefont {Hao}, \citenamefont
  {Feng},\ and\ \citenamefont {Li}}]{Song2019}%
  \BibitemOpen
  \bibfield  {author} {\bibinfo {author} {\bibfnamefont {C.}~\bibnamefont
  {Song}}, \bibinfo {author} {\bibfnamefont {J.}~\bibnamefont {Cui}}, \bibinfo
  {author} {\bibfnamefont {H.}~\bibnamefont {Wang}}, \bibinfo {author}
  {\bibfnamefont {J.}~\bibnamefont {Hao}}, \bibinfo {author} {\bibfnamefont
  {H.}~\bibnamefont {Feng}}, \ and\ \bibinfo {author} {\bibfnamefont
  {Y.}~\bibnamefont {Li}},\ }\href {\doibase 10.1126/sciadv.aaw5686} {\bibfield
   {journal} {\bibinfo  {journal} {Sci. Adv.}\ }\textbf {\bibinfo {volume}
  {5}},\ \bibinfo {pages} {eaaw5686} (\bibinfo {year} {2019})}\BibitemShut
  {NoStop}%
\bibitem [{\citenamefont {Chen}\ \emph {et~al.}(2019)\citenamefont {Chen},
  \citenamefont {Sun},\ and\ \citenamefont {Dong}}]{Chen2019}%
  \BibitemOpen
  \bibfield  {author} {\bibinfo {author} {\bibfnamefont {J.-F.}\ \bibnamefont
  {Chen}}, \bibinfo {author} {\bibfnamefont {C.-P.}\ \bibnamefont {Sun}}, \
  and\ \bibinfo {author} {\bibfnamefont {H.}~\bibnamefont {Dong}},\ }\href
  {\doibase 10.1103/physreve.100.062140} {\bibfield  {journal} {\bibinfo
  {journal} {Phys. Rev. E}\ }\textbf {\bibinfo {volume} {100}},\ \bibinfo
  {pages} {062140} (\bibinfo {year} {2019})}\BibitemShut {NoStop}%
\bibitem [{\citenamefont {Deffner}\ and\ \citenamefont
  {Lutz}(2013)}]{Deffner2013}%
  \BibitemOpen
  \bibfield  {author} {\bibinfo {author} {\bibfnamefont {S.}~\bibnamefont
  {Deffner}}\ and\ \bibinfo {author} {\bibfnamefont {E.}~\bibnamefont {Lutz}},\
  }\href {\doibase 10.1103/physreve.87.022143} {\bibfield  {journal} {\bibinfo
  {journal} {Phys. Rev. E}\ }\textbf {\bibinfo {volume} {87}},\ \bibinfo
  {pages} {022143} (\bibinfo {year} {2013})}\BibitemShut {NoStop}%
\bibitem [{\citenamefont {Crooks}(2007)}]{Crooks2007}%
  \BibitemOpen
  \bibfield  {author} {\bibinfo {author} {\bibfnamefont {G.~E.}\ \bibnamefont
  {Crooks}},\ }\href {\doibase 10.1103/physrevlett.99.100602} {\bibfield
  {journal} {\bibinfo  {journal} {Phys. Rev. Lett.}\ }\textbf {\bibinfo
  {volume} {99}},\ \bibinfo {pages} {100602} (\bibinfo {year}
  {2007})}\BibitemShut {NoStop}%
\bibitem [{\citenamefont {Scandi}\ and\ \citenamefont
  {Perarnau-Llobet}(2019)}]{Scandi2018}%
  \BibitemOpen
  \bibfield  {author} {\bibinfo {author} {\bibfnamefont {M.}~\bibnamefont
  {Scandi}}\ and\ \bibinfo {author} {\bibfnamefont {M.}~\bibnamefont
  {Perarnau-Llobet}},\ }\href {\doibase 10.22331/q-2019-10-24-197} {\bibfield
  {journal} {\bibinfo  {journal} {Quantum}\ }\textbf {\bibinfo {volume} {3}},\
  \bibinfo {pages} {197} (\bibinfo {year} {2019})}\BibitemShut {NoStop}%
\end{thebibliography}%

\end{document}